\documentclass[11pt]{article}
\usepackage{amsmath}
\usepackage{epsfig,bm}
\def\b{\begin}
\def\e{\end}

\def\be{\b{equation}}
\def\ee{\e{equation}}

\def\nt{\noindent}
\def\tr{\mbox{Tr}}
\def\i{{\cal I}}

\begin{document}

\title{COLLAPSE OF THE ENTANGLED STATE AND THE ENTROPY INCREASE IN AN
ISOLATED SYSTEM
\thanks{ Project supported by
the National Nature Science Foundation of China (Grant
No.10305001).}}

\author{Qi-Ren Zhang
\\CCAST(World Lab),P.O.Box 8730,Beijing,100080\\and\\
Department of Technical Physics, Peking University ,
Beijing,100871,China\thanks{Mailing address.}}

\maketitle

\vskip0.3cm

\begin{abstract}
We show that the collapse of the entangled quantum state makes the
entropy increase in an isolated system. The second law of
thermodynamics is thus proven in its most general form.


\end{abstract}

To understand the foundation of the second law of thermodynamics is
a long standing problem in physics. The H-theorem of Boltzmann is a
classical proof of the law. It is based on a model of colliding
classical particle system for the macroscopic matter, therefore is
not general enough, even from the view point of the classical
statistical physics. Here we show an information theoretical proof
of the law. It is based on the state entanglement in time
development and the state collapse in measurement, therefore is
quite general.

The time development of the density operator $\rho(t)$ for an
isolated system is governed by the von Neumann equation. Its
solution is \be \rho(t)=U(t,t_0)\rho(t_0)U(t_0,t)\; ,\label{1}\ee in
which $U(t,t_0)$ is the time displacement operator of the state from
time $t_0$ to time $t$. Defining the information \be {\cal I}(t)=\tr
[\rho(t)\ln\rho(t)]\label{2}\ee at time t, we see from (\ref{1}) \be
{\cal I}(t)=\tr[U(t,t_0)\rho(t_0)\ln\rho(t_0)U(t_0,t)]
=\tr[\rho(t_0)\ln\rho(t_0)U(t_0,t)U(t,t_0)]={\cal I}(t_0)\;
,\label{10}\ee because of $U(t_0,t)U(t,t_0)=1$. It is the
information conservation in quantum mechanics.

To measure the entropy of a system, one has to divide the system
into macroscopically infinitesimal parts. The entropy of the $i$th
part is defined to be $S_i=-k_B\tr(\rho_i\ln\rho_i)$, in which
$\rho_i$ is the reduced density operator of the part $i$. The
entropy of the whole system is defined to be the sum \be S=
\sum_iS_i=-k_B\sum_i\tr(\rho_i\ln\rho_i)\ee of the entropies of
these parts, as an extensive thermodynamical variable should be.
When one measures the entropy of the system at time $t_0$, he has
destroyed the entanglement of the states of various parts of the
system. The state and the density operator of the system are
therefore factorized. Under this condition, the entropy of the
system is \be
S(t_0)=-k_B\sum_i\tr[\rho_i(t_0)\ln\rho_i(t_0)]=-k_B\tr[\rho(t_0)\ln\rho(t_0)]
=-k_B{\cal I}(t_0)\; .\ee For an isolated system, the information
conservation (\ref{10}) works. The information of the system at
$t>t_0$ is therefore \be {\cal I}(t)=-S(t_0)/k_B\; .\label{s1}\ee
During the period from time $t_0$ to $t$, the interaction between
different parts of the system makes their states be entangled again.
It means the states of different parts are correlated. If one
measures the entropy of the system at time $t$, he has to measure
the entropies of every part of the system, and therefore destroy
this entanglement once more. This is the state collapse, and causes
the loss of correlation information. Since the parts of the system
are not isolated, their information is not conserved. It makes the
entropy \be S(t)=-k_B\sum_i\tr[\rho_i(t)\ln\rho_i(t)]\ee at time $t$
does not equal $S(t_0)$ in general. By intuition we see, the sum of
information of all parts of the system should not be more than the
information of the system, since the correlation information of
various parts is not included in the sum. It is \be
\sum_i\tr[\rho_i(t)\ln\rho_i(t)]\leq {\cal I}(t)\; .\label{s2} \ee
If this is true, we obtain \be S(t)\geq S(t_0)\label{s3}\ee from
(\ref{s1})-(\ref{s2}) for an isolated system. This is exactly the
second law of thermodynamics.

To prove the statement (\ref{s2}), let us remind you the following
mathematical inequalities. To make this paper be self-contained, we
also collect the proofs of these inequalities here, although they
may be found in text books.

\nt Lemma 1. For any positive number $x$ we have  \be x\ln x\geq
x-1\; , \label{11}\ee the equality holds when and only when $x=1$.

Proof: It may be verified by differentiation, that $x\ln x-( x-1)$
as a function of positive variable $x$ has unique minimum 0 at
$x=1$. The lemma is therefore proven.

\nt Lemma 2. For set $[w_i]$ of positive numbers and set $[x_i]$ of
non-negative numbers with $\sum_ix_i=1$, we have \be
\sum_ix_iw_i\ln\sum_ix_iw_i\leq\sum_ix_iw_i\ln w_i\; .\label{12}\ee

Proof: The average $\bar{w}\equiv\sum_ix_iw_i$ is positive. By lemma
1 we see
\begin{eqnarray} \sum_ix_iw_i\ln\sum_ix_iw_i-\sum_ix_iw_i\ln
w_i=\sum_ix_iw_i\ln\frac{\bar{w}}{w_i}\nonumber\\=-\sum_ix_i\bar{w}\frac{w_i}{\bar{w}}
\ln\frac{w_i} {\bar{w}}\leq
-\sum_ix_i\bar{w}(\frac{w_i}{\bar{w}}-1)=0\;
.\nonumber\end{eqnarray} The lemma is therefore proven.

\nt Lemma 3. For set $[W_i]$ of positive numbers and set $[T_{ij}]$
of non-negative numbers with \be \sum_iW_i=1\;\;\;\;\;
\mbox{and}\;\;\;\; \; \sum_iT_{ij}=\sum_jT_{ij}=1 \; ,\label{0}\ee
we have \be W_j^\prime \equiv\sum_iW_iT_{ij}>0 \;\;\;\;\mbox{for
every} \;j ,\label{a}\ee\be\sum_jW_j^\prime=1\; ,\label{b}\ee and
\be \sum_jW_j^\prime\ln W_j^\prime\leq\sum_iW_i\ln W_i\;
.\label{13}\ee

Proof: (\ref{a}) and (\ref{b}) are obvious. By (\ref{0}) and lemma 2
we see \be \sum_jW_j^\prime\ln W_j^\prime =\sum_j\left(
\sum_iW_iT_{ij}\right)\ln \left(\sum_iW_iT_{ij}\right)\leq
\sum_{ij}W_iT_{ij}\ln W_i=\sum_i W_i\ln W_i\; .\nonumber \ee  The
lemma is therefore proven.

\nt Lemma 4. For positive numbers $[W_{ij}]$, $W_i=\sum_jW_{ij}$ and
$W_j^\prime=\sum_iW_{ij}$, with $\sum_{ij}W_{ij}=1$, we have \be
\sum_iW_i=1\;,\;\;\;\;\;\;\sum_jW_j^\prime =1\;,\label{14}\ee and
\be \sum_{ij}W_{ij}\ln W_{ij}\geq \sum_iW_i\ln W_i
+\sum_jW_j^\prime\ln W_j^\prime\; .\label{15}\ee The equality holds
when and only when $W_{ij}=W_iW_j^\prime$ for all $ij$, it is that
the $W_{ij}$ may be factorized.

Proof: (\ref{14}) is obvious. By lemma 1 we see \be
\frac{W_{ij}}{W_iW_j^\prime}\ln \frac{W_{ij}}{W_iW_j^\prime}\geq
\frac{W_{ij}}{W_iW_j^\prime}-1\; ,\label{c} \ee the equality holds
when and only when $W_{ij}=W_iW_j^\prime$. Multiplying two sides of
(\ref{c}) by the positive number $W_iW_j^\prime$ and summing up over
$ij$, one obtains \be \sum_{ij}W_{ij}\ln W_{ij}- \sum_iW_i\ln W_i
-\sum_jW_j^\prime\ln W_j^\prime\geq 0\; .\nonumber\ee This is
exactly (\ref{15}). The lemma is therefore proven.

Suppose $[L]$ is a complete set of commutative dynamical variables
of the system, with a complete orthonormal set of eigenstates
$[|n\rangle]$. The $[L]$ representation of density operator $\rho$
is a matrix with elements $\rho_{n,n^\prime}=\langle
n|\rho|n^\prime\rangle$. If $\rho$ itself is included in the set
$[L]$, the $[L]$ representation of $\rho$ is called natural. In a
natural representation, the density matrix is diagonal:
$\rho_{n,n^\prime}=W_n\delta_{n,n^\prime}$, in which $W_n$ is the
$n$th eigenvalue of $\rho$, denoting the probability of finding the
system being in the state $|n\rangle$. The information (\ref{2}) may
be written in the form \be \i =\sum_nW_n\ln W_n\; .\label{d}\ee  One
may also consider the information about a specially chosen complete
set of commutative dynamical variables $[L]$, with complete set of
orthonormal eigenstates $[|m\rangle]$. For an ensemble of the
systems with the density operator $\rho$ , the probability of
finding the system in the state $|m\rangle$ is \be
W^\prime_m=\sum_n\langle m|n\rangle W_n\langle n|m\rangle\;
.\label{e}\ee The definition of the information about the variables
$[L]$ is \be \i_{[L]}\equiv\sum_mW^\prime_m\ln W^\prime_m\; .\ee
Since $|\langle n|m\rangle|^2$ are non-negative, and $\sum_n|\langle
n|m\rangle|^2=\sum_m|\langle n|m\rangle|^2=1$, according to lemma 3
and equation (\ref{d}) we have \be \i_{[L]}\leq\i\; . \ee

Now, let us divide the system into two parts $a$ and $b$. Suppose
$[L_i]$, with $i=a$ or $b$, is a complete set of commutative
dynamical variables of part $i$, $|n_i\rangle$ is their $n_i$th
eigenstate, and $[|n_i\rangle]$ is a complete set of states of part
$i$. Therefore $[|n_an_b\rangle]\equiv[|n_a\rangle|n_b\rangle]$ is a
complete orthonormal set of states of the system. In the $[L_aL_b]$
representation, The density operator of the system is a matrix, with
elements \be \rho_{n_an_b,n^\prime_an^\prime_b}\equiv\langle
n_an_b|\rho|n^\prime_a n^\prime_b\rangle\; .\ee From (\ref{e}) we
see the probability of finding part $a$ in the state $|n_a\rangle$
and part $b$ in the state $|n_b\rangle$ is \be
W_{n_an_b}=\sum_n\langle n_an_b|n\rangle W_n\langle
n|n_an_b\rangle\; ,\ee with normalization \be
\sum_{n_an_b}W_{n_an_b}=1\; .\label{z}\ee The information of
dynamical variables $[L_a,L_b]$ is \be
\i_{L_a,L_b}=\sum_{n_a,n_b}W_{n_an_b}\ln W_{n_an_b}\leq\i\;
.\label{f}\ee The probability of finding part $a$ in the state
$|n_a\rangle$ and the probability of finding the part $b$ in the
state $|n_b\rangle$ are \be
W_{n_a}=\sum_{n_b}W_{n_an_b}\;\;\;\;\mbox{and}\;\;\;\;W^\prime_{n_b}
=\sum_{n_a}W_{n_an_b}\; .\label{g}\ee respectively. In
(\ref{z}-\ref{g}), it is understood that the summation is over those
$n_a$ and $n_b$ only, for which $W_{n_an_b}>0$.

The density operator $\rho_a$ of part $a$ is reduced from the
density operator $\rho$ of the system. In the representation
$[L_a]$, it is a matrix with elements \be
(\rho_a)_{n_a,n^\prime_a}=\sum_{n_b}\rho_{n_an_b,n^\prime_an_b}
=\sum_{n_b}\langle n_an_b|\rho |n^\prime_an_b\rangle\; ,\ee and may
be written in a compact form \be \rho_a=\tr_b\rho \; .\ee The
subscript $b$ denotes that the trace is a sum of matrix elements
diagonal with respect to quantum numbers of part $b$ only. Likewise,
$\rho_b=\tr_a\rho $. Suppose $\rho_i$ is included in the set
$[L_i]$, the probability of finding the part $i$ in state
$|n_i\rangle$ is its eigenvalue $W_{n_i}$, and is expressed in
(\ref{g}). The information about part $i$ is \be
\i_i=\tr\rho_i\ln\rho_i=\sum_{n_i}W_{n_i}\ln W_{n_i}\; .\ee From
lemma 4 and equations (\ref{f},\ref{g}) we see,\be
\tr\rho_a\ln\rho_a+\tr\rho_b\ln\rho_b\leq\tr\rho\ln\rho\;
,\label{h}\ee the equality holds when and only when the density
operator of the system may be factorized into a direct product of
density operators of its parts. We may further subdivide the parts
and apply (\ref{h}) to them again and again, the result is the
statement (\ref{s2}). This statement, together with the second law
(\ref{s3}) of thermodynamics, is therefore finally proven.

The proof here is quite general. It seems relying on the quantum
mechanical effects of state entanglement and its collapse. However,
it is still more general. It is an information theoretical proof,
relies only on the information conservation (\ref{10}) and the
general relations (\ref{g}) of the probabilities. The former is a
character of dynamics. But it is shared by quantum dynamics and
classical dynamics, as well as some dynamics not yet have been
discovered at the present time. The later is purely mathematical.
State entanglement and its collapse is only a special way for their
realization. They may be realized in classical mechanics or in some
unknown mechanics as well. The second law of thermodynamics is
therefore almost dynamics independent, except the requirement of
information conservation. It may be still exactly true in the
future, even though one day people find that the quantum mechanics
is only approximate. It is also quite generally applicable, not only
to thermodynamics but also to any statistical science, including
social science, if the information conservation is true for them.
From the proof we learn that the entropy of an isolated system
increases only because one loses the correlation information between
different parts of the system. It opens a possibility of developing
a theory which takes the correlation information into account.
\end{document}